\documentclass[aps,prl,preprint,floatfix]{revtex4}

\usepackage{amsmath}
\usepackage{array}
\usepackage{supertabular}
\usepackage{mathtext}

\begin{document}

\title{On the opportunity of spectroscopic determination of absolute
atomic densities in non-equilibrium plasmas from measured relative
intensities within resonance multiplets distorted by
self-absorption.}
\author{B. P. ~Lavrov}
\author{A. S. ~Mikhailov}
\address{Faculty of Physics, St.-Petersburg State University, St.-Petersburg, 198904, Russia\\
E-mail address: lavrov@pobox.spbu.ru}

\date{\today}

\begin{abstract}
The opportunities of the application of the recently proposed
approach in optical emission spectroscopy of non-equilibrium
plasmas have been studied. The approach consists of several
methods of the determination of {\em absolute} particle densities
of atoms from measured {\em relative} intensities within resonance
multiplets distorted by self-absorption. All available
spectroscopic data concerning resonance spectral lines of atoms
having multiplet ground states from boron up to gallium were
analyzed. It is found that in the case of C, O, F, S and Cl atoms
an application of the methods needs VUV technique, while densities
of B, Al, Si, Sc, Ti, V, Co, Ni, Ga atoms may be obtained by means
of the intensity measurements in UV and visible parts of emission
spectra suitable for ordinary spectrometers used for optical
diagnostics and monitoring of non-equilibrium plasmas including
industrial plasma technologies.
\end{abstract}

\maketitle

The opportunities of the application of the recently proposed
approach [1---3] in optical emission spectroscopy of
non-equilibrium plasmas have been studied. The approach consists
of several methods of the determination of absolute particle
densities of atoms from measured relative intensities within
resonance multiplets distorted by self-absorption. The main idea
of the approach is that in the case of sufficient self-absorption
for a pair of resonance lines having common upper level the ratio
of intensities doesn't depend on the population density of upper
level, but depends on the length of homogeneous plasma column
along optical axis L, the gas temperature T, and populations of
the sublevels of multiplet structure of the ground state. In the
case of Boltzmann distribution over the sublevels with the
temperature T, the intensity ratio depends on the temperature T
and the product NL, where N is the total population of the ground
state, usually close to the atomic particle density.

All available spectroscopic data concerning resonance spectral
lines of atoms having multiplet ground states from Boron up to
Gallium were analyzed (see Table 1). It is found that in the case
of the C, O, F, S and Cl atoms an application of the methods needs
VUV technique, while densities of the B, Al, Si, Sc, Ti, V, Co, Ni
and Ga atoms may be obtained by means of the intensity
measurements in UV and visible parts of emission spectrum suitable
for ordinary middle class spectrometers used for optical
diagnostics and monitoring of non-equilibrium plasmas including
industrial plasma technologies. All spectroscopic data needed for
experimental determination of particle density of the B, Al, Si,
Sc, Ti, V, Co, Ni and Ga atoms in ionized gases and  plasma
 are presented in the Table 2, where the  $\lambda_{ik}$ -- is the wavelength of a spectral line in Angstrems,
 the $\nu_{ki}$ - its wavenumber in cm$^{-1}$ ,
the $E_i$ is energy of lower level of the line in cm$^{-1}$, the
$g_k$  and $g_i$ are degeneracies of upper and lower levels, and
 the $A_{ki}$ is spontaneous emission transition probability (Einstein coefficient) in 10$^6s^{-1}$.

The data collected in the Table 2 may be also used for computer
modelling calculations of the intensity ratios as functions of the
optical thickness NL and the gas temperature T (determining line
broadening) under certain assumptions on the population density
distribution over ground state sublevels. In the case of Boltzmann
distribution corresponding to the gas temperature such
calculations give the opportunity to obtain certain three
dimensional nomogramms (the intensity ratio against NL and T)
useful for {\em a priory} analysis of an applicability of the
method in certain experimental conditions like in [3], as well as
for simple, graphical determination of the atomic densities from
measured relative line intensities like in [1].

\bigskip

The present work was supported by Russian Foundation for Basic
Research (RFBR), Grant № 06-03-32663a.

\newpage

  \tablecaption{ The list of
first 17 elements of the Mendeleev Periodic Table, having
multiplet structure of the ground states(1 - the chemical symbols
of atoms, 2 - the configurations of outer electrons in the ground
state, 3 - the spectroscopic serial symbols of the multiplet
structure sublevels , 4 - the sublevel energies $E_i$ (cm$^{-1}$)).
} \tabletail{\hline} \label{Elem}

\tablehead{\hline
1 & 2 & 3 & 4 & 1 & 2 & 3 & 4  \\
\hline}
\begin{center}
\begin{supertabular}{|c|c|c|c||c|c|c|c|}
\hline
B & $2s^22p$ & $^2P_{1/2}$ & 0       & Sc  & $3d4s^2$   & $^2D_{3/2}$ & 0\\
   &          & $^2P_{3/2}$ & 15,254 &     &             & $^2D_{5/2}$ & 168,34\\
\hline
 C & $2s^22p$ & $^3P_0$     & 0      & Ti  & $3d^24s^2$ & $a^3F_2$ & 0 \\
   &          & $^3P_1$ & 16,4       &     &             & $a^3F_3$ & 170,132\\
   &          & $^3P_1$ & 43,4       &     &              & $a^4F_{9/2}$ & 552,96\\
\hline
 O & $2s_22p^4$ & $^3P_2$ & 0        & V   & $3d^34s^2$ & $a^4F_{3/2}$ & 0\\
   &            & $^3P_1$ & 158,265  &     &            & $a^4F_{5/2}$ & 137.38\\
   &            & $^3P_0$ & 226,977  &     &            & $a^4F_{7/2}$ & 323,46\\

\cline{1-4}
 F & $2s^22p^5$ & $^2P_{3/2}$ & 0    &     &            & $a^4F_{9/2}$ & 552,96\\
\cline{5-8}
   &            & $^2P_{1/2}$ & 404,1 & Co & $3p^63d^74s^2$ & $a^4F_{9/2}$ & 0\\

\cline{1-4}
 Al & $3s^23p$ & $^2P_{1/2}$ & 0      &    &             & $a^4F_{7/2}$ & 816,0\\
    &          & $^2P_{3/2}$ & 112,061&    &             & $a^4F_{5/2}$ & 1406,84\\

\cline{1-4}
 Si & $3s^23p^2$ & $^3P_0$ &  0      &     &             & $a^4F_{3/2}$ & 1809,33\\
 \cline{5-8}
    &            & $^3P_1$ & 77,115   & Ni & $3d^8(3F)4s^2$ & $^3F_4$ & 0\\
    &            & $^3P_2$ & 223,157  &    &             & $^3F_3$ & 1332,164\\

\cline{1-4}
S   & $3s^23p^4$ & $3P_2$ &           &    &             & $^3F_2$ & 2216,550\\
\cline{5-8}
    &            & $3P_1$ & 396,055   & Ga & $4s^24p$    & $^2P_{1/2}$ & 0\\
    &            & $3P_0$ & 573.640   &     &            & $^2P_{3/2}$ & 826,19\\
\cline{1-4}
Cl  & $3s^23p^5$ & $^2P_{3/2}$ &   0    &     &            &      &        \\
    &            & $^2P_{3/2}$&882,3515 &      &           &       &        \\
\hline

\end{supertabular}
\end{center}

\newpage
\tablecaption{The list of resonance lines of B, Al,
Si, Sc, Ti, V, Co, Ni, Ga atoms prospective for spectroscopic
diagnostics of non-equilibrium plasma.}
{\footnotesize
\tablehead{\hline
the transitions & $\lambda_{ik}$, \AA~  & $\nu_{ki}$, cm$^{-1}$ & $E_i$, cm$^{-1}$ & $g_k$ & $g_i$ & $A_{ki}$, 10$^6s^{-1}$ \\

\hline} \tabletail{\hline}

\label{Line}
\begin{supertabular*}{167mm}{@{\extracolsep{\fill}}|c|c|c|c|c|c|c|}
\hline

\multicolumn{7}{|c|}{{\bf B}}\\
\hline
 $3s~^2S_{1/2}\to2p~^2P_{3/2}$ & 2497,723  & 40024,40  &  15,254 & 2 & 4 & 240 \\
 $3s~^2S_{1/2}\to2p~^2P_{1/2}$ & 2496,771  & 40039,65  &   0     & 2 & 2 & 120 \\
\hline \hline
\multicolumn{7}{|c|}{{\bf Al}}\\
\hline
    $8s~^2S_{1/2}\to3p~^2P_{1/2}$ & 2199,15 & 45457 & 0 & 2 & 2 & 1,75\\
    $8s~^2S_{1/2}\to3p~^2P_{3/2}$ & 2204,59 & 45345,183 & 112,061 & 2 & 4 & 3,49\\

    $6d~^2D_{5/2}\to3p~^2P_{1/2}$ &{\em 2204,660} & 45344 & 0 & 4 & 2 & 45,3 \\

    $7s~^2S_{1/2}\to3p~^2P_{1/2}$ & 2257,999 & 44273 & 0 & 2 & 2 & 3,77\\
    $7s~^2S_{1/2}\to3p~^2P_{3/2}$ & 2263,731 & 44161,072 & 112,061 & 2 & 4 & 7,5\\

    $5d~^2D_{3/2}\to3p~^2P_{1/2}$ & 2263,462 & 44166 & 0 & 4 & 2 & 66\\
    $5d~^2D_{3/2}\to3p~^2P_{3/2}$ & 2269,22  & 44054,337 & 112,061& 4 & 4 & 13\\

    $4d~^2D_{3/2}\to3p~^2P_{1/2}$ & 2367,052 & 42234 & 0 & 4 & 2 & 72\\
    $4d~^2D_{3/2}\to3p~^2P_{3/2}$ & 2373,349 & 42121,687 & 112,061 & 4 & 4 & 14\\

    $4d~^2D_{5/2}\to3p~^2P_{3/2}$ &{\em 2373,124} & 42125,722 & 112,061 & 6 & 4 & 86 \\

    $6s~^2S_{1/2}\to3p~^2P_{1/2}$ & 2372,07 & 42144 & 0 & 2 & 2 & 4,78\\
    $6s~^2S_{1/2}\to3p~^2P_{3/2}$ & 2378,368 & 42032,35 & 112,061 & 2 & 4 & 9,5\\

    $4s~^2S_{1/2}\to3p~^2P_{3/2}$ & 3961,5201 & 25235,695 & 112,061 & 2 & 4 &  98  \\
    $4s~^2S_{1/2}\to3p~^2P_{1/2}$ & 3944,0060 & 25347,756 &   0     & 2 & 2 &  49,3 \\

    $3d~^2D_{3/2}\to3p~^2P_{3/2}$ & 3092,8369 & 32323,392 & 112,061 & 4 & 4 &  12 \\
    $3d~^2D_{5/2}\to3p~^2P_{3/2}$ & 3092,7084 & 32324,735 & 112,061 & 6 & 4 &  74 \\
    $3d~^2D_{3/2}\to3p~^2P_{1/2}$ & 3082,1510 & 32435,453 &   0    & 4 & 2 &  63 \\

    $5s~^2S_{1/2}\to3p~^2P_{3/2}$ & 2660,3863 & 37577,346 & 112,061 & 2 & 4 &  26,4 \\
    $5s~^2S_{1/2}\to3p~^2P_{1/2}$ & 2652,4758 & 37689,407 &   0     & 2 & 2 &  13,3 \\

    $y~^2D_{3/2}\to3p~^2P_{3/2}$ & 2575,3962 & 38817,352 & 112,061 & 4 & 4 &   4,4 \\
    $y~^2D_{5/2}\to3p~^2P_{3/2}$ &{\em 2575,0940} & 38821,907 & 112,061 & 6 & 4 &  28     \\
    $y~^2D_{3/2}\to3p~^2P_{1/2}$ & 2567,9823 & 38929,413 &   0     & 4 & 2 &  23     \\

\hline \hline
\multicolumn{7}{|c|}{{\bf Si}}\\
\hline
    $3p^3~^3D_1\to3p^2~^3P_0$ & 2207,98 & 45276 & 0 & 3 & 1 & 31,1\\
    $3p^3~^3D_1\to3p^2~^3P_1$   & 2211,74 & 45199,073 & 77,115 & 3 & 3 & 23,2\\
    $3p^3~^3D_1\to3p^2~^3P_2$   & 2218,91 & 45053,031 & 223,157 & 3 & 5 & 1,5\\

    $3p^3~^3D_2\to3p^2~^3P_1$   & 2210,89 & 45216,514 & 77,115 & 5 & 3  & 41,6\\
    $3p^3~^3D_2\to3p^2~^3P_2$   & 2218,06 & 45070,472 & 223,157 & 5 & 5 & 13,8\\

    $3p4s~^3P_{1}\to3p^2~^3P_{2}$ & 2528,508  & 39537,128 & 223,157 & 3 & 5 &  77     \\
    $3p4s~^3P_{0}\to3p^2~^3P_{1}$ & 2524,108  & 39606,046 &  77,115 & 1 & 3 & 181     \\
    $3p4s~^3P_{1}\to3p^2~^3P_{1}$ & 2519,202  & 39683,170 &  77,115 & 3 & 3 &  45,6 \\
    $3p4s~^3P_{2}\to3p^2~^3P_{2}$ & 2516,112  & 39731,896 & 223,157 & 5 & 5 & 121     \\
    $3p4s~^3P_{1}\to3p^2~^3P_{0}$ & 2514,316  & 39760,285 &   0     & 3 & 1 &  61     \\
    $3p4s~^3P_{2}\to3p^2~^3P_{1}$ & 2506,897  & 39877,938 &  77,115 & 5 & 3 &  46,6 \\

    $3p4s~^1P_{1}\to3p^2~^3P_{2}$ & 2452,118  & 40768,731 & 223,157 & 3 & 5 &   0,60 \\
    $3p4s~^1P_{1}\to3p^2~^3P_{1}$ & 2443,364  & 40914,773 &  77,115 & 3 & 3 &   0,69 \\
    $3p4s~^1P_{1}\to3p^2~^3P_{0}$ & 2438,768  & 40991,888 &   0     & 3 & 1 &   0,74 \\
\hline \hline

\multicolumn{7}{|c|}{{\bf Sc}}\\
\hline

      $3d4s(^3D)4p~^4F_{3/2} \to 3d4s^2~2D_{3/2}$ & 6378,83 & 15673 & 0 & 4 & 4 & 0,16\\
      $3d4s(^3D)4p~^4F_{3/2} \to 3d4s^2~2D_{5/2}$ & 6448,08 & 15504 & 168,34 & 4 & 6 & 0,026\\
      $3d4s(^3D)4p~^4F_{5/2} \to 3d4s^2~2D_{3/2}$ & 6344,83 & 15757 & 0 & 6 & 4& 0.024\\
      $3d4s(^3D)4p~^4F_{5/2} \to 3d4s^2~2D_{5/2}$ & 6413,35 & 15588,23 & 168,34 & 6 & 6 & 0,13\\

      $3d4s(^3D)4p~^4D_{3/2} \to 3d4s^2~^2D_{3/2}$ & 6239,8 & 16022 & 0 & 4 & 4 & 0,71\\
      $3d4s(^3D)4p~^4D_{3/2} \to 3d4s^2~^2D_{5/2}$ & 6305,99 & 15853,48 & 168,34 & 4 &6 & 0,087\\
      $3d4s(^3D)4p~^4D_{5/2} \to 3d4s^2~^2D_{3/2}$ & 6193,7  & 16141 & 0 & 6 & 4 & 0.036 \\
      $3d^2(^3F)4p~^2F_{5/2}\to 3d^2(^1D)4s~^2D_{5/2}$& {\em6193,7} &16141,03 & 17012,76 & 6 & 6 & - \\ 
      $3d4s(^3D)4p~^4D_{5/2} \to 3d4s^2~^2D_{5/2}$ & 6258,9 & 15972,72 & 168,34 & 6 & 6 & 0,45\\

      $3d4s(^3D)4p~^2D_{5/2} \to 3d4s^2~^2D_{3/2}$ & 6239,44 & 16023 & 0 & 6 & 4 & 0,152 \\
      $3d4s(^3D)4p~^2D_{5/2} \to 3d4s^2~^2D_{5/2}$ & 6305,65 & 15854,39 & 168,34 & 6 & 6 & 1,61 \\
      $3d4s(^3D)4p~^2D_{3/2} \to 3d4s^2~^2D_{3/2}$ & 6210,66 & 16097 & 0 & 4 & 4 & 1,28\\
      $3d4s(^3D)4p~^2D_{3/2} \to 3d4s^2~^2D_{5/2}$ & 6276,28 & 15928,56 & 168,34 & 4 & 6 & 0,105\\

      $3d4s(^1D)4p~^2F_{5/2} \to 3d4s^2~^2D_{3/2}$& 4753,165 & 21033 & 0 & 6 & 4 & 1,08\\
      $3d4s(^1D)4p~^2F_{5/2} \to 3d4s^2~^2D_{5/2}$& 4791,52 & 20864,41 & 168,34 & 6 & 6 & 0,197\\

      $3d4s(^3D)4p~^2D_{3/2} \to 3d4s^2~^2D_{3/2}$& 4020,387 & 24866 & 0 & 4 & 4 & 163\\
      $3d4s(^3D)4p~^2D_{3/2} \to 3d4s^2~^2D_{5/2}$& 4047,797 & 24697,83& 168,34& 4& 6 & 15,4\\

      $3d4s(^3D)4p~^2F_{5/2} \to 3d4s^2~^2D_{3/2}$& 3907,484 & 25585 & 0 & 6 &4 & 166\\
      $3d4s(^3D)4p~^2F_{5/2} \to 3d4s^2~^2D_{5/2}$& 3933,375 & 25416,3 & 168,34 & 6 & 6& 16,2\\

      $3d4s^2(^3D)4p~^2P_{3/2}\to3d4s^2~^2D_{3/2}$ & 3255,676 & 30707&      0 & 4 & 4 &32\\
      $3d4s^2(^3D)4p~^2P_{3/2}\to3d4s^2~^2D_{5/2}$ & 3273,628 & 30538,32 & 168,34 & 4 & 6 &281\\

      $3d4s^2(^3D)4p~^2D_{5/2}\to3d4s^2~^2D_{3/2}$ & 3996,601 & 25014 & 0 & 6 & 4 & 16,5\\
      $3d4s^2(^3D)4p~^2D_{5/2}\to3d4s^2~^2D_{5/2}$ & 4023,678 & 24845,87 & 168,34 & 6 & 6 & 165\\


      $3d^2(^3F)4p~^2F_{5/2}\to3d4s^2~^2D_{3/2}$ & 3015,36 & 33154 & 0 & 6 & 4 & 78\\
      $3d^2(^3F)4p~^2F_{5/2}\to3d4s^2~^2D_{5/2}$ & 3030,759 & 32985,45 & 168,34 & 6 & 6 & 10\\
\hline \hline
\multicolumn{7}{|c|}{{\bf Ti}}\\
\hline
      $3s^3(a^2D)4p~s^3D_2\to3d^24s^2~a^3F_2$                  & 2519,04 & 39686     & 0        & 5 & 5 & 5,9   \\
      $3s^3(a^2D)4p~s^3D_2\to3d^24s^2~a^3F_3$                  & 2529,85 & 39515.97  & 170,132  & 5 & 7 & 38    \\

      $3s^3(a^2D)4p~s^3D_3\to3d^24s^2~a^3F_3$                  & 2527,98 & 39545,305 & 170,132  & 7 & 7 & 6,8   \\
      $3s^3(a^2D)4p~s^3D_3\to3d^24s^2~a^3F_4$                  & 2541,92 & 39328,563 & 386,874  & 7 & 9 & 43    \\

      $3d^3(^2G)4p~t^3F_4\to3d^24s^2~a^3F_3$                   & 2596,58 & 38500,578 & 170,132  & 9 & 7 & 6,9   \\
      $3d^3(^2G)4p~t^3F_4\to3d^24s^2~a^3F_4$                   & 2611,28 & 38283,836 & 386,874  & 9 & 9 & 64    \\

      $3d^3(^2G)4p~t^3F_2\to3d^24s^2~a^3F_2$                   & 2599,92 & 38451     & 0        & 5 & 5 & 67    \\
      $3d^3(^2G)4p~t^3F_2\to3d^24s^2~a^3F_3$                   & 2611,48 & 38281,166 & 170,132  & 5 & 7 & 33    \\

      $3d^3(^2G)4g~t^3F_3\to3d^24s^2~a^3F_2$                   & 2593,64 & 38544     & 0        & 7 & 5 & 6,9   \\
      $3d^3(^2G)4g~t^3F_3\to3d^24s^2~a^3F_3$                   & 2605,15 & 38374,25  & 170,132  & 7 & 7 & 64    \\
      $3d^3(^2G)4g~t^3F_3\to3d^24s^2~a^3F_4$                   & 2619,94 & 38157,51  & 386,874  & 7 & 9 & 21    \\

      $3d^2(^3P)4s4p(^1P)~u^3D_3\to3d^24s^2~a^3F_3$            & 2631,54 & 37989,58  & 170,132  & 7 & 7 & 17    \\
      $3d^2(^3P)4s4p(^1P)~u^3D_3\to3d^24s^2~a^3F_4$            & 2646,64 & 37772,84  & 386,874  & 7 & 9 & 150    \\

      $3d^2(^3P)4s4p(^1P)~u^3D_2\to3d^24s^2~a^3F_2$            & 2632,42 & 37977     & 0        & 5 & 5 & 27    \\
      $3d^2(^3P)4s4p(^1P)~u^3D_2\to3d^24s^2~a^3F_3$            & 2644,26 & 37806,65  & 170,132  & 5 & 7 & 140    \\

      $3d^2(^1G)4s4p(^3P)~v^3F_3\to3d^24s^2~a^3F_2$            & 2933,55 & 34079     & 0        & 7 & 5 & 9,6  \\
      $3d^2(^1G)4s4p(^3P)~v^3F_3\to3d^24s^2~a^3F_3$            & 2948,26 & 33908,448 & 170,132  & 7 & 7 & 93   \\
      $3d^2(^1G)4s4p(^3P)~v^3F_3\to3d^24s^2~a^3F_4$            & 2967,22 & 33691,706 & 386,874  & 7 & 9 & 11    \\

      $3d^2(^1G)4s4p(^3P)~v^3F_4\to3d^24s^2~a^3F_3$            & 2937,32 & 34034,839 & 170,132  & 9 & 7 & 7,7    \\
      $3d^2(^1G)4s4p(^3P)~v^3F_4\to3d^24s^2~a^3F_4$            & 2956,13 & 33818,097 & 386,874  & 9 & 9 & 97      \\

      $3d^2(^1G)4s4p(^3P)~v^3F_2\to3d^24s^2~a^3F_2$            & 2942    & 33981     & 0        & 5 & 5 & 100     \\
      $3d^2(^1G)4s4p(^3P)~v^3F_2\to3d^24s^2~a^3F_3$            & 2956,8  & 33810,507 & 170,132  & 5 & 7 & 18      \\

      $3d^3(^4F)4p~w^3G_3\to3d^24s^2~a^3F_2$                   & 3186,45 & 31374     & 0        & 7 & 5 & 80     \\
      $3d^3(^4F)4p~w^3G_3\to3d^24s^2~a^3F_3$                   & 3203,83 & 31203,669 & 170,132  & 7 & 7 & 7,2    \\

      $3d^3(^4F)4p~w^3G_4\to3d^24s^2~a^3F_3$                   & 3191,99 & 31319,319 & 170,132  & 9 & 7 & 85     \\
      $3d^3(^4F)4p~w^3G_4\to3d^24s^2~a^3F_4$                   & 3214,24 & 31102,577 & 386,874  & 9 & 9 & 6,5    \\

      $3d^24s^2(^1G)4s4p(^3P)~x^3G_3\to3d^24s^2~a^3F_2$        & 3341,88 & 29915     & 0        & 7 & 5 & 65     \\
      $3d^24s^2(^1G)4s4p(^3P)~x^3G_3\to3d^24s^2~a^3F_4$        & 3385,66 & 29527,846 & 386,874  & 7 & 9 & 5,2    \\

      $3d^24s^2(^1G)4s4p(^3P)~x^3G_4\to3d^24s^2~a^3F_3$        & 3354,64 & 29800,946 & 170,132  & 9 & 7 & 69    \\
      $3d^24s^2(^1G)4s4p(^3P)~x^3G_4\to3d^24s^2~a^3F_4$        & 3379,22 & 29584,204 & 386,874  & 9 & 9 & 6,2    \\

      $3d^2(^3F)4s4p(^1P)~w^3D_2\to3d^24s^2~a^3F_2$            & 3358,28 & 29769     & 0        & 5 & 5 & 7,6    \\
      $3d^2(^3F)4s4p(^1P)~w^3D_2\to3d^24s^2~a^3F_3$            & 3377,48 & 29598,523 & 170,132  & 5 & 7 & 69    \\

      $3d^2(^3F)4s4p(^1P)~y^3G_3\to3d^24s^2~a^3F_2$            & 3635,46 & 27499     & 0        & 7 & 5 & 80,4   \\
      $3d^2(^1G)4s4p(^3p)~v^3F_3\to3d^24s^2~a^3F_3$            & 3658,1 & 27328,843  & 170,132  & 7 & 7 & 5,83   \\
      $3d^2(^1G)4s4p(^3p)~v^3F_3\to3d^24s^2~a^3F_4$            & 3687,35 & 27112,101 & 386,874  & 7 & 9 & 0,35   \\

      $3d^2(^3F)4s4p(^1P)~y^3G_4\to3d^24s^2~a^3F_3$            & 3642,68 & 27444,535 & 170,132  & 9 & 7 & 77,4   \\
      $3d^2(^3F)4s4p(^1P)~y^3G_4\to3d^24s^2~a^3F_4$            & 3671,67 & 27227,793 & 386,874  & 9 & 9 & 4,59    \\

      $3d^2(^1D)4s4p(^3P)~x^3D_2\to3d^24s^2~a^3F_2$            & 3646,2  & 27418     & 0        & 5 & 5 & 2,6    \\
      $3d^2(^1D)4s4p(^3P)~x^3D_2\to3d^24s^2~a^3F_3$            & 3668,97 & 27247     & 170,132  & 5 & 7 & 5,4    \\

      $3d^2(^1D)4s4p(^3P)~x^3D_3\to3d^24s^2~a^3F_2$            & 3637,97 & 27480     & 0        & 7 & 5 & 0,93   \\
      $3d^2(^1D)4s4p(^3P)~x^3D_3\to3d^24s^2~a^3F_3$            & 3660,63 & 27309,915 & 170,132  & 7 & 7 & 3    \\
      $3d^2(^1D)4s4p(^3P)~x^3D_3\to3d^24s^2~a^3F_4$            & 3689,91 & 27093,173 & 386,874  & 7 & 9 & 3,53   \\

      $3d^2(^1D)4s4p(^3P)~x^3F_4\to3d^24s^2~a^3F_3$            & 3722,57 & 26855,52  & 170,132  & 9 & 7 & 3,4\\
      $3d^2(^1D)4s4p(^3P)~x^3F_4\to3d^24s^2~a^3F_4$            & 3752,86 & 26638,778 & 386,874  & 9 & 9 & 50,4\\

      $3d^2(^1D)4s4p(^3P)~x^3F_2\to3d^24s^2~a^3F_2$            & 3729,82 & 26803     & 0        & 5 & 5 & 42,7\\
      $3d^2(^1D)4s4p(^3P)~x^3F_2\to3d^24s^2~a^3F_3$            & 3753,64 & 26633,285 & 170,132  & 5 & 7 & 8,2\\

      $3d^2(^3F)4s4p(^1P)~y^3F_3\to3d^24s^2~a^3F_2$            & 3962,85 & 25227     & 0        & 7 & 5 & 4,13\\
      $3d^2(^3F)4s4p(^1P)~y^3F_3\to3d^24s^2~a^3F_3$            & 3989,76 & 25057,085 & 170,132  & 7 & 7 & 37,9\\
      $3d^2(^3F)4s4p(^1P)~y^3F_3\to3d^24s^2~a^3F_4$            & 4024,57 & 24840,343 & 386,874  & 7 & 9 & 6,14\\

      $3d^2(^3P)4s4p(^3P)~y^5D_3\to3d^24s^2~a^3F_3$            & 3900,96 & 25627,47  & 170,132  & 7 & 7 & 1,28\\
      $3d^2(^3P)4s4p(^3P)~y^5D_3\to3d^24s^2~a^3F_4$            & 3934,24 & 25410,73  & 386,874  & 7 & 9 & 0,45\\

      $3d^2(^3P)4s4p(^3P)~z^3P_2\to3d^24s^2~a^3F_2$            & 3921,42 & 25494     & 0        & 5  & 5 & 2,15\\
      $3d^2(^3P)4s4p(^3P)~z^3P_2\to3d^24s^2~a^3F_3$            & 3947,78 & 25323,59  & 170,132  & 5  & 7 & 9,6 \\

      $3d^2(^3F)4s4p(^1P)~y^3F_3\to3d^24s^2~a^3F_2$            & 3962,85 & 25227     & 0        & 7   & 5 & 4,13\\
      $3d^2(^3F)4s4p(^1P)~y^3F_3\to3d^24s^2~a^3F_3$            & 3989,76 & 25057,085 & 170,132  & 7  & 7 & 37,9\\
      $3d^2(^3F)4s4p(^1P)~y^3F_3\to3d^24s^2~a^3F_4$            & 4024,57 & 24840,343 & 386,874  & 7 & 9 & 6,14\\

      $3d^2(^3F)4s4p(^1P)~y^3F_4\to3d^24s^2~a^3F_3$            & 3964,27 & 25218,202 & 170,132  & 9 & 7 & 3,09\\
      $3d^2(^3F)4s4p(^1P)~y^3F_4\to3d^24s^2~a^3F_4$            & 3998,64 & 25001,46  & 386,874  & 9 & 9 & 40,8\\

      $3d^2(^3F)4s4p(^1P)~y^3F_2\to3d^24s^2~a^3F_2$            & 3981,76 & 25107     & 0        & 5 & 5 & 37,6\\
      $3d^2(^3F)4s4p(^1P)~y^3F_2\to3d^24s^2~a^3F_3$            & 4008,93 & 24937,285 & 170,132  & 5 & 7 & 7,03\\

      $3d^2(^3P)4s4p(^3P)~z^5S_2\to3d^24s^2~a^3F_2$            & 3982,48 & 25103     &  0       & 5 & 5 & 4,5\\
      $3d^2(^3P)4s4p(^3P)~z^5S_2\to3d^24s^2~a^3F_3$            & 4009,66 & 24932,75  & 170,132  & 5 & 7 & 1,21\\

      $3d^2(^3F)4s4p(^3P)~z^3G_3\to3d^24s^2~a^3F_2$            & 4656,47 & 21469     &  0       & 7 & 5 & 1,99\\
      $3d^2(^3F)4s4p(^3P)~z^3G_3\to3d^24s^2~a^3F_3$            & 4697,68 & 21299,362 & 170,132  & 7 & 7 & 0,085\\

      $3d^2(^3F)4s4p(^3P)~z^3G_4\to3d^24s^2~a^3F_3$            & 4667,59 & 21418,364 & 170,132  & 9 & 7 & 2,18\\
      $3d^2(^3F)4s4p(^3P)~z^3G_4\to3d^24s^2~a^3F_4$            & 4715,3  & 21201,622 & 386,874  & 9 & 9 & 0,069\\

      $3d^2(^3F)4s4p(^3P)~z^3D_2\to3d^24s^2~a^3F_2$            & 4997,10  & 20006    & 0        & 5 & 5 & 0,47\\
      $3d^2(^3F)4s4p(^3P)~z^3D_2\to3d^24s^2~a^3F_3$            & 5039,95  & 19835,900& 170,132  & 5 & 7 & 3,89\\

      $3d^2(^3F)4s4p(^3P)~z^3D_3\to3d^24s^2~a^3F_3$            & 5009,65  & 19955,923& 170,132  & 7 & 7 & 0,209\\
      $3d^2(^3F)4s4p(^3P)~z^3D_3\to3d^24s^2~a^3F_4$            & 5064,66  & 19739,181& 386,874  & 7 & 9 & 3,79\\

      $3d^2(^3F)4s4p(^3P)~z^3F_3\to3d^24s^2~a^3F_2$             & 5147,48 & 19422     & 0       & 7 & 5 & 0,35\\
      $3d^2(^3F)4s4p(^3P)~z^3F_3\to3d^24s^2~a^3F_3$            & 5192,98 & 19251,444 & 170,132 & 7 & 7 & 3,49\\
      $3d^2(^3F)4s4p(^3P)~z^3F_3\to3d^24s^2~a^3F_4$            & 5252,11 & 19034,702 & 386,874 & 7 & 9 & 0,123\\

      $3d^2(^3F)4s4p(^3P)~z^3F_4\to3d^24s^2~a^3F_3$            & 5152,20 & 19403,836 & 170,132 & 9 & 7 & 0,264\\
      $3d^2(^3F)4s4p(^3P)~z^3F_4\to3d^24s^2~a^3F_4$            & 5210,39 & 19187,094 & 386,874 & 9 & 9 & 3,57\\

      $3d^2(^3F)4s4p(^3P)~z^3F_2\to3d^24s^2~a^3F_2$            & 5173,75 & 19323     & 0       & 5 & 5 & 3,8\\
      $3d^2(^3F)4s4p(^3P)~z^3F_2\to3d^24s^2~a^3F_3$            & 5219,71 & 19152,856 & 170,132 & 5 & 7 & 0,25\\

      $3d^2(^3F)4s4p(^3P)~z^5D_3\to3d^24s^2~a^3F_3$            & 5426,26 & 18423,86  & 170,132 & 7 & 7 & 0,0319\\
      $3d^2(^3F)4s4p(^3P)~z^5D_3\to3d^24s^2~a^3F_4$            & 5490,84 & 18207,12  & 386,874 & 7 & 9 & 0,014\\

 \hline
 \hline

\multicolumn{7}{|c|}{{\bf V}}\\
\hline
     $3d^3(^4F)4s4p(^1P)~w^4F_{7/2}\to3d^34s^2~a^4F{5/2}$      & 3043,12 & 32851,46  & 137,38 & 8 & 6 & 23\\
     $3d^3(^4F)4s4p(^1P)~w^4F_{7/2}\to3d^34s^2~a^4F{7/2}$      & 3060,46 & 32665,38  & 323,46 & 8 & 8 & 140\\
     $3d^3(^4F)4s4p(^1P)~w^4F_{7/2}\to3d^34s^2~a^4F{9/2}$      & 3082,11 & 32435,88  & 552,96 & 8 &10 &21\\

     $3d^3(^4F)4s4p(^1P)~w^4F_{5/2}\to3d^34s^2~a^4F{3/2}$ & 3043,56 & 32847 & 0 & 6 & 4 & 18\\
     $3d^3(^4F)4s4p(^1P)~w^4F_{5/2}\to3d^34s^2~a^4F{5/2}$ & 3056,33 & 32709,44 & 137,38& 6 & 6 & 130\\

     $3d^3(^4F)4s4p(^1P)~w^4F_{9/2}\to3d^34s^2~a^4F{7/2}$ & 3044,94 & 32831,84 & 323,46 & 10 & 8 & 12\\
     $3d^3(^4F)4s4p(^1P)~w^4F_{9/2}\to3d^34s^2~a^4F{9/2}$ & 3066,38 & 32602,34 & 552,96 & 10 & 10 & 210\\

     $3d^3(^4F)4s4p(^1P)~x^4G_{7/2}\to3d^34s^2~a^4F{5/2}$ & 3183,41 & 31403,77 & 137,38 & 8 & 6 & 240\\
     $3d^3(^4F)4s4p(^1P)~x^4G_{7/2}\to3d^34s^2~a^4F{7/2}$ & 3202,38 & 31217,69 & 323,46 & 8 & 8 & 40\\

     $3d^3(^4F)4s4p(^1P)~x^4G_{9/2}\to3d^34s^2~a^4F{7/2}$ & 3183,98 & 31398,25 & 323,46 & 10 & 8 & 250\\
     $3d^3(^4F)4s4p(^1P)~x^4G_{9/2}\to3d^34s^2~a^4F{9/2}$ & 3207,41 & 31168,78 & 552,96 & 10 & 10 & 26\\

     $3d^4(^5D)4p~y^4D_{3/2}\to3d^34s^2~a^4F{3/2}$        & 3808,52 & 26249    & 0   & 4 & 4 & 14,8\\
     $3d^4(^5D)4p~y^4D_{3/2}\to3d^34s^2~a^4F{5/2}$        & 3828,56 & 26112,1  & 137,38 & 4 & 6 & 53,3\\

     $3d^4(^5D)4p~y^4D_{5/2}\to3d^34s^2~a^4F{5/2}$        & 3813,49 & 26215,27 & 137,38 & 6 & 6 & 11,9\\
     $3d^4(^5D)4p~y^4D_{5/2}\to3d^34s^2~a^4F{7/2}$        & 3840,75 & 26029,19 & 323,46 & 6 & 8 & 54,8\\

     $3d^4(^5D)4p~y^4D_{7/2}\to3d^34s^2~a^4F{7/2}$        & 3822,01 & 26156,83 & 323,46 & 8 & 8 & 8,1\\
     $3d^4(^5D)4p~y^4D_{7/2}\to3d^34s^2~a^4F{9/2}$        & 3855,37 & 25927,33 & 552,96 & 8 & 10& 57,8\\

     $3d^3(^4F)4s4p(^3P)~z^2G_{9/2}\to3d^34s^2~a^4F{7/2}$  & 3841,89  & 26021,44 & 323,46 & 10 & 8 & 0,95\\
     $3d^3(^4F)4s4p(^3P)~z^2G_{9/2}\to3d^34s^2~a^4F{9/2}$  & 3876,09  & 25791,94 & 552,96 & 10 & 10 & 14,8\\

     $3d^4(^5D)4p~y^4F_{5/2}\to3d^34s^2~a^4F_{3/2}$        & 3844,44 & 26004     & 0 & 6 & 4 & 6\\
     $3d^4(^5D)4p~y^4F_{5/2}\to3d^34s^2~a^4F_{5/2}$        & 3864,86 & 25866,85  & 137,38 & 6 & 6 & 27\\
     $3d^4(^5D)4p~y^4F_{5/2}\to3d^34s^2~a^4F_{7/2}$        & 3892,86 & 25680,77  & 323,46 & 6 & 8 & 8,2\\

     $3d^4(^5D)4p~y^4F_ {7/2}to3d^34s^2~a^4F_{5/2}$        & 3847,33 & 25984,7   & 137,38 & 8 & 6 & 4,8\\
     $3d^4(^5D)4p~y^4F_ {7/2}to3d^34s^2~a^4F_{7/2}$        & 3875,08 & 25798,62  & 323,46 & 8 & 8 & 23,6\\
     $3d^4(^5D)4p~y^4F_ {7/2}to3d^34s^2~a^4F_{9/2}$        & 3909,89 & 25569,12  & 552,96 & 8 & 10 & 4,3\\

     $3d^4(^5D)4p~y^4F_{3/2}\to3d^34s^2~a^4F_{3/2}$        & 3855,37 & 25931     & 0 & 4 & 4 & 33\\
     $3d^4(^5D)4p~y^4F_{3/2}\to3d^34s^2~a^4F_{5/2}$        & 3875,9  & 25793,17  & 137,38 & 4 & 6 & 8,3\\

     $3d^3(^4F)4s4p(^3P)~z^2G_(7/2)\to3d^34s^2~a^4F_{5/2}$ & 3862,22 & 25884,54  & 137,38 & 8 & 6 & 1,13\\
     $3d^3(^4F)4s4p(^3P)~z^2G_(7/2)\to3d^34s^2~a^4F_{7/2}$ & 3890,18 & 25698,46  & 323,46 & 8 & 8 & 6,6\\
     $3d^3(^4F)4s4p(^3P)~z^2G_(7/2)\to3d^34s^2~a^4F_{9/2}$ & 3925,24 & 25468,96  & 552,96 & 8 & 10 & 0,97\\

     $3d^4(^5D)4p~y^4F_{9/2}\to3d^34s^2~a^4F_{7/2}$        & 3867,6  & 25848,46  & 323,46 & 10 & 8 & 2,54\\
     $3d^4(^5D)4p~y^4F_{9/2}\to3d^34s^2~a^4F_{9/2}$        & 3902,25 & 25618,96  & 552,96 & 10 & 10 & 26,8\\

     $3d^3(^4F)4s4p(^3P)~z^4F_{7/2}\to3d^34s^2~a^4F_{5/2}$ & 4306,21 & 23215,72 & 137,38 & 8 & 6 & 1,2\\
     $3d^3(^4F)4s4p(^3P)~z^4F_{7/2}\to3d^34s^2~a^4F_{7/2}$ & 4341,01 & 23029,64 & 323,46 & 8 & 8 & 5 \\

     $3d^3(^4F)4s4p(^3P)~z^2G_(5/2)\to3d^34s^2~a^4F_{3/2}$ & 4307,18 & 23211    & 0 & 6 & 4 & 1,11\\
     $3d^3(^4F)4s4p(^3P)~z^2G_(5/2)\to3d^34s^2~a^4F_{5/2}$ & 4332,82 & 23073,16 & 137,38 & 6 & 6 & 4,6\\
     $3d^3(^4F)4s4p(^3P)~z^2G_(5/2)\to3d^34s^2~a^4F_{7/2}$ & 4368,04 & 22887,08 & 323,46 & 6 & 8 & 1,01\\

     $3d^3(^4F)4s4p(^3P)~z^4F_{9/2}\to3d^34s^2~a^4F_{9/2}$ & 4352,87 & 22966,91 & 552,96 & 10 & 10 & 5,8 \\
     $3d^3(^4F)4s4p(^3P)~z^4F_{9/2}\to3d^34s^2~a^4F_{7/2}$ & 4309,8  & 23196,41 & 323,46 & 10 & 8  & 0,94\\

     $3d^3(^4F)4s4p(^3P)~z^4F_{3/2}\to3d^34s^2~a^4F_{3/2}$ & 4330,02 & 23088    & 0 & 4 & 4 & 5,2\\
     $3d^3(^4F)4s4p(^3P)~z^4F_{3/2}\to3d^34s^2~a^4F_{5/2}$ & 4355,94 & 22950    & 137,38 & 4 & 6 & 1,05\\

     $3d^3(^4F)4s4p(^3P)~z^4G_{5/2}\to3d^34s^2~a^4F_{3/2}$ & 4577,17 & 21841    & 0      & 6 & 4 & 4,75\\
     $3d^3(^4F)4s4p(^3P)~z^4G_{5/2}\to3d^34s^2~a^4F_{5/2}$ & 4606,15 & 21704,04 & 137,38 & 6 & 6 & 0,69 \\

     $3d^3(^4F)4s4p(^3P)~z^4G_{7/2}\to3d^34s^2~a^4F_{5/2}$ & 4580,4  & 21826,07 & 137,38 & 8 & 6 & 4,71\\
     $3d^3(^4F)4s4p(^3P)~z^4G_{7/2}\to3d^34s^2~a^4F_{7/2}$ & 4619,77 & 21639,99 & 323,46 & 8 & 8 & 0,62\\
     $3d^3(^4F)4s4p(^3P)~z^4G_{9/2}\to3d^34s^2~a^4F_{7/2}$ & 4586,36 & 21797,61 & 323,46 & 10 & 8 & 5,1\\
     $3d^3(^4F)4s4p(^3P)~z^4G_{9/2}\to3d^34s^2~a^4F_{9/2}$ & 4635,18 & 21568,11  & 552,96 & 10 & 10 &0,37\\

     $3d^3(^4F)4s4p(^3P)~z^4D_{5/2}\to3d^34s^2~a^4F_{3/2}$ & 4799,77 & 20828 &    0 & 6 & 4 & 0,127\\
     $3d^3(^4F)4s4p(^3P)~z^4D_{5/2}\to3d^34s^2~a^4F_{5/2}$ & 4831,64 & 20691,1 &  137,38 & 6 & 6 & 2\\
     $3d^3(^4F)4s4p(^3P)~z^4D_{5/2}\to3d^34s^2~a^4F_{7/2}$ & 4875,48 & 20505,02 & 323,46 &6 & 8 & 7,3\\

     $3d^3(^4F)4s4p(^3P)~z^4D_{7/2}\to3d^34s^2~a^4F_{7/2}$ & 4827,45 & 20709,05 & 323,46 & 8 & 8 & 1,19\\
     $3d^3(^4F)4s4p(^3P)~z^4D_{7/2}\to3d^34s^2~a^4F_{9/2}$ & 4881,56 & 20479,55 & 552,96 & 8 & 10 & 7,7\\

     $3d^3(^4F)4s4p(^3P)~z^4D_{3/2}\to3d^34s^2~a^4F_{3/2}$ & 4832,43 & 20688 & 0 & 4 & 4 & 2,23\\
     $3d^3(^4F)4s4p(^3P)~z^4D_{3/2}\to3d^34s^2~a^4F_{5/2}$ & 4864,74 & 20550,38 & 137,38 &4 & 6 & 7,7\\
    \hline
    \hline

    \multicolumn{7}{|c|}{{\bf Co}}\\
\hline
     $3p^63d^7(^2G)4s4p(^3P)~w^4F_{9/2}\to3p^63d^74s^2~a^4F_{9/2}$ & 2309,02 & 43295 & 0 & 10 & 10 & 56\\
     $3p^63d^7(^2G)4s4p(^3P)~w^4F_{9/2}\to3p^63d^74s^2~a^4F_{7/2}$ & 2353,42 & 42479,32 & 816 & 10 & 8 & 15\\

     $3p^63d^7(^2G)4s4p(^1P)~x^4F_{7/2}\to3p^63d^74s^2~a^4F_{9/2}$ & 2384,86 & 41918 & 0 & 8 &10 & 24\\
     $3p^63d^7(^2G)4s4p(^1P)~x^4F_{7/2}\to3p^63d^74s^2~a^4F_{7/2}$ & 2432,21 & 41102,41 & 816 & 8 & 8 & 260\\

     $3p^63d^7(^2G)4s4p(^1P)~x^4F_{5/2}\to3p^63d^74s^2~a^4F_{7/2}$ & 2402,06  & 41618,23 & 816 & 6 & 8 & 51\\
     $3p^63d^7(^2G)4s4p(^1P)~x^4F_{5/2}\to3p^63d^74s^2~a^4F_{5/2}$ & 2436,66  & 41027,39 & 1406,84 & 6 & 6 & 260\\

     $3p^63d^7(^2G)4s4p(^1P)~x^4F_{7/2}\to3p^63d^74s^2~a^4F_{9/2}$ & 2521,36 & 39649 & 0 & 8 & 10 & 300\\
     $3p^63d^7(^2G)4s4p(^1P)~x^4F_{7/2}\to3p^63d^74s^2~a^4F_{7/2}$ & 2574,35 & 38833,16 & 816 & 8 & 8 & 17\\

     $3p^63d^7(^2G)4s4p(^1P)~x^4F_{5/2}\to3p^63d^74s^2~a^4F_{7/2}$ & 2528,97 & 39529,95 & 816 & 6 & 8 & 280\\
     $3p^63d^7(^2G)4s4p(^1P)~x^4F_{5/2}\to3p^63d^74s^2~a^4F_{5/2}$ & 2567,35 & 38939,11 & 1406,84 & 6 & 6 & 30\\

     $3p^63d^7(^2G)4s4p(^1P)~x^4D_{3/2}\to3p^63d^74s^2~a^4F_{5/2}$ & 2535,96 & 39420,93 & 1406,84 & 4 & 6 & 190\\
     $3p^63d^7(^2G)4s4p(^1P)~x^4D_{3/2}\to3p^63d^74s^2~a^4F_{3/2}$ & 2562,15 & 39018,44 & 1809,33 & 4 & 4 & 39\\

     $3p^63d^7(^4F)4s4p(^3P)~z^4D_{7/2}\to3p^63d^7(^2G)4s4p(^1P)~x^4F_{9/2}$ & 3412,63 & 29295 & 0 & 8 & 10 & 12\\
     $3p^63d^7(^4F)4s4p(^3P)~z^4D_{7/2}\to3p^63d^7(^2G)4s4p(^1P)~x^4F_{7/2}$ & 3510,43 & 28478,52 & 816 & 8 & 8 & 3,8\\

     $3p^63d^7(^4F)4s4p(^3P)~z^4D_{5/2}\to3p^63d^7(^2G)4s4p(^1P)~x^4F_{7/2}$ & 3520,08 & 28400,37 & 816 & 6 & 8 & 4,6\\
     $3p^63d^7(^4F)4s4p(^3P)~z^4D_{5/2}\to3p^63d^7(^2G)4s4p(^1P)~x^4F_{5/2}$ & 3594,87 & 27809,53 & 1406,84 &6 & 6 & 9,2\\
\hline \hline

\multicolumn{7}{|c|}{{ Ni}}\\
\hline

     $3d^8(^3F)4s4p(^3P)~^3D_3\to3d^8(^3F)4s^2~^3F_4$ &  2984,13 & 33501 & 0 & 7 & 9 & 6,6\\
     $3d^8(^3F)4s4p(^3P)~^3D_3\to3d^8(^3F)4s^2~^3F_2$ &  3195,57 & 31284,272 & 2216,550 & 7 & 5 & 0,58\\

     $3d^8(^3F)4s4p(^3P)~^3F_3\to3d^8(^3F)4s^2~^3F_3$ & 3145,72 & 31780,170 & 1332,164 & 7 & 7 & 0,8\\
     $3d^8(^3F)4s4p(^3P)~^3F_3\to3d^8(^3F)4s^2~^3F_4$ & 3019,14 & 33112 & 0  & 7 & 9 &6,4\\

     $3d^8(^3F)4s4p(^3P)~^3D_2\to3d^8(^3F)4s^2~^3F_3$ & 3097,12 & 32278 & 1332,164 & 5 & 7 & 3,3\\
     $3d^8(^3F)4s4p(^3P)~^3D_2\to3d^8(^3F)4s^2~^3F_2$ & 3184,37 & 31394,340 & 2216,55 & 5 & 5 & 0,71\\

     $3d^9(^3F)4s^2~^1F_3\to3d^8(^3F)4s^2~^3F_4$ & 3221,65 & 31031 & 0 & 7 & 9 & 1,6\\
     $3d^9(^3F)4s^2~^1F_3\to3d^8(^3F)4s^2~^3F_3$ & 3366,17 & 29698 & 1332,164 & 7 & 7 & 4\\
     $3d^9(^3F)4s^2~^1F_3\to3d^8(^3F)4s^2~^3F_2$ & 3469,49 & 28814,47 & 2216,55 & 7 & 5 & 1,3\\

     $3d^8(^3F)4s4p(^3P)~^3G_3\to3d^8(^3F)4s^2~^3F_3$ & 3282,7 & 30453,998 & 1332,164 & 7 & 7 & 0,6\\
     $3d^8(^3F)4s4p(^3P)~^3G_3\to3d^8(^3F)4s^2~^3F_2$ & 3380,85 & 29569,612 & 2216,55 & 7 & 5 & 3,8\\

     $3d^9(^2D)4p~^3F_4\to3d^8(^3F)4s^2~^3F_4$ & 3391,05 & 29481 & 0 & 9 & 9 & 6,6\\
     $3d^9(^2D)4p~^3F_4\to3d^8(^3F)4s^2~^3F_4$ & 3551,53 & 28148,825 & 1332,164 & 9 & 7 & 0,16\\

     $3d^9(^3F)4s^2~^3F_3\to3d^8(^3F)4s^2~^3F_4$ & 3409,58 & 29321 & 0 & 7 & 9 & 0,37\\
     $3d^9(^3F)4s^2~^3F_3\to3d^8(^3F)4s^2~^3F_3$ & 3571,87 & 27988,598 & 1332,164 & 7 & 7 & 5,2\\
     $3d^9(^3F)4s^2~^3F_3\to3d^8(^3F)4s^2~^3F_2$ & 3688,42 & 27104,212 & 2216,550 & 7 & 5 & 0,45\\

     $3d^9(^3F)4s^2~^3F_2\to3d^8(^3F)4s^2~^3F_3$ & 3413,48 & 29287,25 & 1332,164 & 5 & 7 & 3,8\\
     $3d^9(^3F)4s^2~^3F_2\to3d^8(^3F)4s^2~^3F_2$ & 3519,77 & 28402,864 & 2216,550 & 5 & 5 & 4,1\\

     $3d^9(^2D)4p~^3D_2\to3d^8(^3F)4s^2~^3F_3$ & 3500,85 & 28556,313 & 1332,164 & 5 & 7 & 4,6\\
     $3d^9(^2D)4p~^3D_2\to3d^8(^3F)4s^2~^3F_2$ & 3612,74 & 27671,927 & 2216,550 & 5 & 5 & 4,2\\

     $3d^8(^3F)4s4p(^3P)~5G_4\to3d^8(^3F)4s^2~^3F_4$ & 3561,75 & 28068 & 0 & 9 & 9 & 0,29\\
     $3d^8(^3F)4s4p(^3P)~5G_4\to3d^8(^3F)4s^2~^3F_3$ & 3739,23 & 26735,901 &1332,164 & 9 & 7 & 0,24\\

\hline \hline

\multicolumn{7}{|c|}{{\bf Ga}}\\
\hline
    $5s~^2S_{1/2}\to4p~^2P_{3/2}$ & 4172,04   & 23962,34  & 826,19  & 2 & 4 &  92 \\
    $5s~^2S_{1/2}\to4p~^2P_{1/2}$ & 4032,99   & 24788,53  &   0     & 2 & 2 &  49 \\

    $4d~^2D_{3/2}\to4p~^2P_{3/2}$ & 2944,17   & 33955,47  & 826,19  & 4 & 4 &  27 \\
    $4d~^2D_{5/2}\to4p~^2P_{3/2}$ & 2943,64   & 33961,66  & 826,19  & 6 & 4 & 140 \\
    $4d~^2D_{3/2}\to4p~^2P_{1/2}$ & 2874,24   & 34781,66  &   0     & 4 & 2 & 120 \\

    $6s~^2S_{1/2}\to4p~^2P_{3/2}$ & 2719,66   & 36758,58  & 826,19  & 2 & 4 &  23 \\
    $6s~^2S_{1/2}\to4p~^2P_{1/2}$ & 2659,87   & 37584,77  &   0     & 2 & 2 &  12 \\
\hline
\end{supertabular*}}
\clearpage
\newpage

\end{document}